\documentclass[10pt]{article}
\usepackage{setspace}

\usepackage{amsmath}

\usepackage{amssymb}

\usepackage{graphicx}

\usepackage{cite}

\usepackage{color} 

\usepackage{graphics,epsfig,amsfonts,amsmath,amssymb}
\usepackage{placeins}
\usepackage{epsfig}
\usepackage{color}

\usepackage[labelfont=bf,labelsep=period,justification=raggedright]{caption}

\makeatletter
\renewcommand{\@biblabel}[1]{\quad#1.}
\makeatother

\date{}
\begin{document}%

\begin{flushleft}
{\Large
\textbf{Investigating the development of chemotherapeutic drug resistance in cancer:  A multiscale computational study\footnote{This paper is a preprint of a paper submitted to IET Systems Biology and is subject to Institution of Engineering and Technology Copyright. If accepted, the copy of record will be available at IET Digital Library}}
}\\
Gibin G Powathil$^{1,\ast}$, 
Mark AJ Chaplain$^{1}$
Maciej Swat$^{2}$
\\
$^{1}$Division of Mathematics, University of Dundee, Dundee, Scotland, DD1 4HN.\\
$^{2}$The Biocomplexity Institute and Department of Physics, Indiana University Bloomington, Bloomington, Indiana, USA.\\

$^{\ast}$E-mail: gibin@maths.dundee.ac.uk
\end{flushleft}

\section*{Abstract}
Chemotherapy is one of the most important therapeutic options used to treat human cancers, either alone or in combination with radiation therapy and surgery. Over the past several years, the concept of combination chemotherapy i.e. combining multiple chemotherapeutic drugs to increase the cytotoxic effect and cell kill rate, has been used in cancer treatment. However, the development of drug resistance by cancer cells continues to be a major impediment to the successful delivery of these multi-drug therapies. Recent studies have indicated that intra-tumoural heterogeneity has a significant role in driving resistance to chemotherapy in many human malignancies. Multiple factors including the internal cell-cycle dynamics and the external microenvironement contribute to the intra-tumoural heterogeneity. In this paper we present a hybrid, multiscale, individual-based mathematical model, incorporating internal cell-cycle dynamics and changes in oxygen concentration, to study the effects of delivery of several different chemotherapeutic drugs on the heterogeneous subpopulations of cancer cells with varying cell-cycle dynamics. The computational simulation results from the multiscale model are in good agreement with available experimental data and support the hypothesis that slow-cycling sub-populations of tumour cells within a growing tumour mass can induce drug resistance to chemotherapy and thus the use of conventional chemotherapy may actually result in the emergence of dominant, therapy-resistant, slow-cycling subpopulations of tumour cells. Our results indicate that the appearance of this chemotherapeutic resistance is mainly due to the inability of the administered drug to target all cancer cells irrespective of the stage in the cell-cycle they are in i.e. most chemotherapeutic drugs target cells in a particular phase/phases of the cell-cycle, and hence always spare some cancer cells that are not in the targeted cell-cycle phase/phases. The results also suggest that this cell-cycle-mediated drug resistance may be overcome by using multiple doses of cell-cycle, phase-specific chemotherapy that targets cells in all phases and its appropriate sequencing and scheduling.

\section*{Introduction}

Anticancer drugs are effective in controlling tumour growth by inflicting damage to various target molecules and thereby triggering multiple cellular and intercellular pathways, leading to cell death or cell-cycle arrest. One of the major impediments in chemotherapy treatment is drug resistance which develops through multiple mechanisms, including multi-drug and cell-cycle mediated resistances to chemotherapy drugs \cite{Bailar1997, Shah2001}. The presence of drug resistance to multiple types of drugs that are available for the treatment of cancers is an indication of a dynamically changing tumour tissue and its microenvironemnt \cite{Saunders2012}. Several recent experimental studies have indicated the fundamental role of intra-tumoural heterogeneity as a driving source for the resistance to multiple chemotherapeutic drugs \cite{Ding2012, Navin2011, Koshkin2012}. These studies imply that cancer cells within tumours are heterogenous due to genotypical and phenotypical reasons and hence their sensitivity to the chemotherapeutic drugs vary accordingly. Genotypical and phenotypical heterogeneities can be a result of multiple factors such as heritable genetic variations, potential mutant cells, stroma-tumour cell interactions, tumour cell/tumour cell interactions and the stochasticity inherent in most biological processes \cite{Saunders2012}. 

Cell-cycle-mediated chemotherapeutic drug resistance is mainly defined as the lack of sensitivity of a cancer cell to a chemotherapeutic drug due to its relative position in the cell-cycle \cite{Shah2001}. The cell-cycle regulates complex processes such as proliferation, cellular division and DNA replication and is partially controlled by a complex hierarchy of metabolic and genetic networks \cite{Schwartz2005}. The complete cell-cycle mechanism can be considered in four different phases depending on key cell-cycle-regulated protein concentrations and DNA synthesis \cite{Douglas2003}. Further, depending on the surrounding microenvironment and intracellular dynamics, cells may sometimes exit from the cell-cycle and enter a quiescence phase or resting phase \cite{Schwartz2005}. Cell-cycle dynamics are also affected by several intracellular and extracellular factors such as Cdk inhibitors that can act as negative regulators of the cell-cycle, nutrient supply, cell size, temperature  and cellular oxygen concentration \cite{Schwartz2005, Goda2003}.  Moreover, the cell-cycle length of each cancer cell varies within the tumour mass, from 10 min per cycle to 50 hr per cycle, further complicating cell-cycle-mediated drug insensitivity  \cite{Douglas2003}. Cell-cycle-mediated drug-resistance and poor performance can be addressed to some extent by an appropriate combination of  cytotoxic drugs that target cells in the various phases of their cell-cycle \cite{Schwartz2005}. As the complex interactions of several intracellular and extracellular factors together with varying cell-cycle dynamics clearly affect a cellÕs response to therapy, it is important to study carefully the effects of cell-cycle phase-specific chemotherapeutic drugs in addressing cell-cycle-mediated drug-resistance. It is in this respect that computational modelling can be especially helpful.

Together with experimental and clinical research, several mathematical and computational modelling approaches have been developed to study the occurrence of drug resistance. In the early 1980s Goldie and Coldman proposed a hypothesis that predicts the occurrence of drug-resistant sub-clones at a rate related to the genetic instability and developed several mathematical models to explain drug-resistance in cancer treatments \cite{Goldie1979, Goldie1982, Goldie1983, Goldie1985, Coldman1985}. Furthermore, the details of several other mathematical models that study drug-resistance can be found in the review by Lavi et al. \cite{Lavi2012} (and the references therein). These approaches help to understand and to some extent analytically quantify various biological processes. It can also be used as a tool to analyse and design drug development experiments and clinical trials. Moreover, these mathematical models serve as test bases to study the responses of chemotherapeutic drugs as well as other therapies towards various perturbations in the cell-cycle kinetics and the changing microenvironment, and related drug resistances. Cancer progression is a complex, multiscale phenomenon and hence it necessitates a multi-scale modelling approach to produce truly predictive mathematical models and to study the responses of multiple treatments to several intra- and inter-cellular fluctuations. A number of hybrid discrete-continuum mathematical models currently exist that account for intracellular and/or microenvironmental heterogeneities to study the responses of multi-modality treatment scenarios. Previously, hybrid modelling has been used extensively to model various aspects of tumour development and progression, such as the formation of multicellular spheroids \cite{Kansal2000, Patel2001}, tumour-induced angiogenesis \cite{Anderson1998}, cancer invasion \cite{Turner2002}. Hybrid modelling approaches have also been used by a number of authors to study the evolution and interaction of a growing tumour in response to changes in the surrounding microenvironment \cite{Ferreira1999, Gerlee2007, Dormann, Ribba2004, Patel2001, Alarcon2003}. There have also been modelling attempts to address the multiscale nature of cellular growth by incorporating a number of details such as vascular dynamics, oxygen transport, hypoxia, cell division and various other intracellular interactions to study the multiscale nature of tumour dynamics  and the response of treatments \cite{Alarcon2005, Mallet2006, Owen2009, Macklin2009, Perfahl2011, Ribba2006b, Zhang2009}. A review of recent models in this area may be found in the paper of Deisboeck et al. \cite{Deisboeck2011}. There have also been efforts to integrate mathematical models of cancer with experimental data in a genuine attempt to develop quantitative, predictive models \cite{Byrne2010}. Together with these developments in the field of cancer modelling, more and more modelling strategies have been developed to study chemotherapy \cite{Agur2006, fister2000, Frieboes2009, liu2007, Mistry2008, panetta1995, Ribba2005} and radiotherapy \cite{Enderling2006, Enderling2007, Ribba2006, richard2007}. Recently, Powathil et al. \cite{Powathil2012b, Powathil2013} developed a hybrid multiscale cellular automaton model, incorporating the dynamical changes during the cell-cycle and oxygen distribution to study cell-cycle-based chemotherapy delivery in combination with radiation therapy. It was shown that an appropriate combination of cell-cycle-specific chemotherapeutic drugs along with radiation delivery could effectively be used to control tumour progression. An overview of the previously developed hybrid multiscale techniques can be found in \cite{Powathil2014}.

In this paper, we present a hybrid multiscale model using a CompuCell3D framework, incorporating the spatio-temporal dynamics at the cell level and the macroscopic changes of tissue oxygen dynamics. This hybrid multiscale model is then used to study the effects of cell-cycle-based chemotherapeutic drugs on cancer cell populations with drug resistance. Here, we consider several different cancer cell populations with varying drug resistances induced by the heterogeneity in cell-cycle dynamics and cell-cycle duration.

 %%%%
\begin{figure}[t!]
  \begin{center}
      \includegraphics[scale=.45]{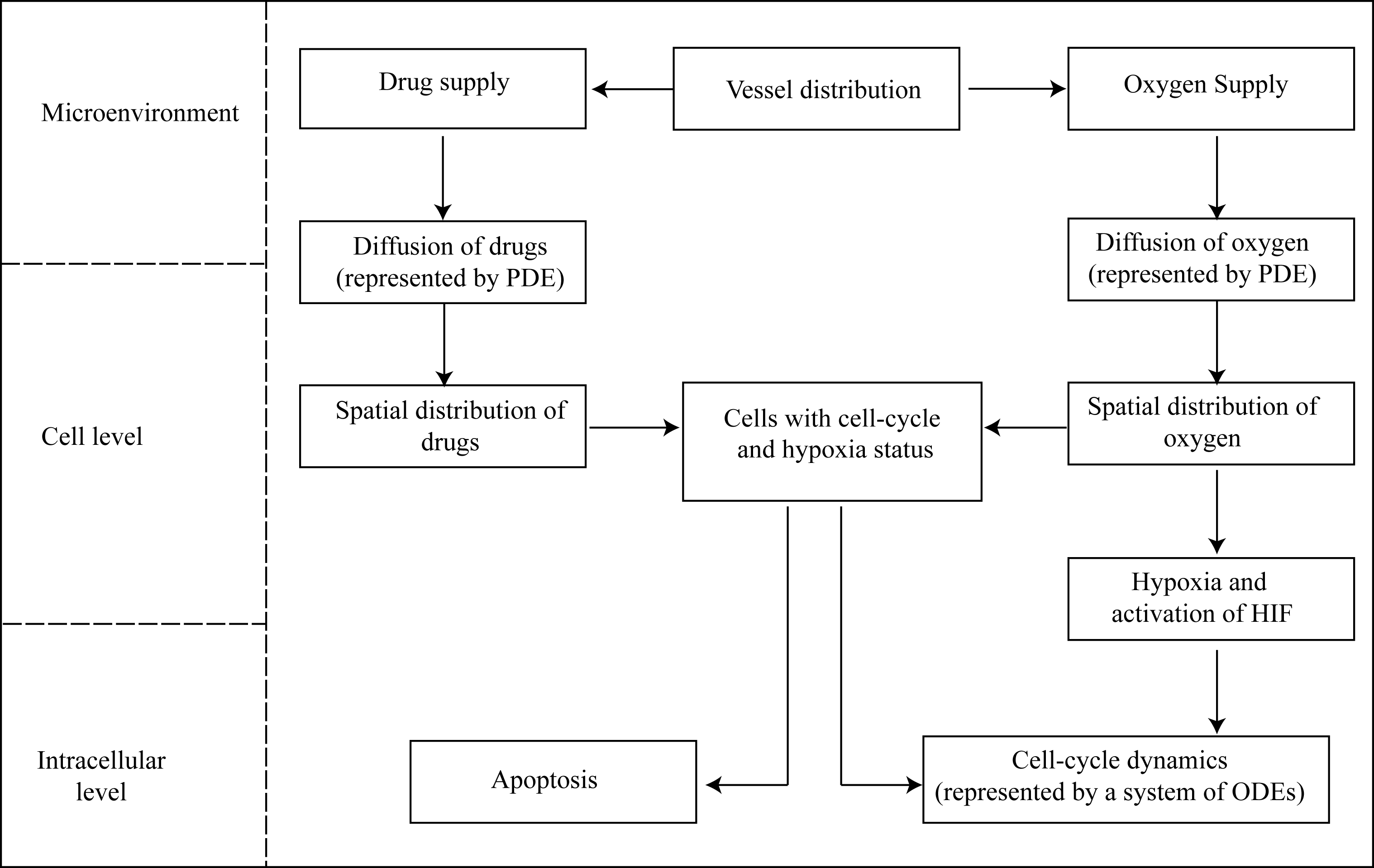}
      \end{center}
  \caption{Schematic diagram showing the various processes and multiple scales that are incorporated into the model (Adapted from Powathil et al. \cite{Powathil2012b})}
  \label{diagram}
\end{figure}
%%%

\section*{Methods and Model}
In order to capture the key dynamics of tumour progression in a mathematical model, it is necessary to couple processes that are occurring at various spatial and temporal scales using multiscale mathematical models. The incorporation of the key complex processes involved in cancer development and progression will further help in devising patient-specific treatment protocols in multimodality cancer therapy. The underlying modelling formulation of a multiscale hybrid-model focussed at the level of individual cancer cells has previously been developed by Powathil et al. {\cite{Powathil2012b}}, where the dynamics of individual cells are modelled using a cellular automaton approach. In this paper we use an alternative modelling approach to study how drug resistance induced by the slow-cycling subpopulation of the cancer cells affects the overall effectiveness of chemotherapy. The cell-level details are considered to model intracellular and intercellular processes, such as cellular response to tissue hypoxia, cell-cycle dynamics and changes in cell-cell adhesion, and the microenvironmental dynamics are modelled using a system of partial differential equations \cite{Powathil2012b, Powathil2013}. The macroscale dynamics are linked to the sub-cellular and cellular changes through a cellular potts modelling approach using the CompuCell3D framework \cite{Andasari2012}. The computational domain contains four different components required to simulate the multiscale model. These are: (1) cancer cells whose spatio-temporal progression is controlled by internal cell-cycle dynamics and the external microenvironment; (2) oxygen distribution; (3) cross-sections of blood vessels from where the oxygen and chemotherapeutic drugs are supplied and (4) concentration of the chemotherapeutic drugs. A schematic diagram showing the various processes that are incorporated into the model and the multiple scales that are involved is given in Figure \ref{diagram}. Other specific details and the mathematical formulation of the multiscale model are given below. 

\subsection*{Cancer growth: The CompuCell3D framework}
To implement our models we used CompuCell3D (CC3D) \cite{Swat2012} (see also http://www.compucell3d.org for full details) -- a modelling framework for building, running, testing and analysing multi-cell, multi-scale, single-cell-based models of tissues. In its current implementation, CC3D models use the Cellular Potts Model (CPM) formalism. Because CPM represents cells as spatially extended domains and focuses on modelling single cells and cellular processes such as adhesion, growth, mitosis, death, chemical secretion and absorption, etc., it is very well-suited to modelling cancer cells and their interactions that evolve with respect to time and space \cite{Andasari2012}.

To simulate the growth of a tumour monolayer from a single cell we use a 2-dimensional lattice of size $300 \times 300$ pixels in the $x-$ and $y-$directions (corresponding to $1\times1$ mm$^2$ of cancer tissue). The initial configuration is one single cancer cell surrounded by a number of blood vessels. Division of the cancer cells and hence growth of the tumour mass is driven by internal cell-cycle dynamics. To simulate the cell-cycle dynamics we associated  a cell-cycle model (implemented as a set of ordinary differential equations ) with each individual cancer cell (details in the next sub-section). We represented the cell-cycle ODE using System Biology Markup Language (SBML) and used BionetSolver to solve the ODE for each cell. We used CC3D Python scripting to implement the dependence of cellular properties on the particular cell-cycle state of a given cell \cite{Andasari2012}. The BionetSolver is a CC3D module that permits easy integration of sophisticated reaction-kinetic models with cell-based models. Internally, BionetSolver uses the SBML ODE Solver Library (soslib - http://www.tbi.univie.ac.at/~raim/odeSolver/) and so numerically integrates SBML models (usually a set of coupled ODEs). To simulate the dependence of the cell-cycle on oxygen and chemotherapy drug concentrations we introduce two diffusive fields (Oxygen and Drug). All CPM models use Monte Carlo Step (MCS) as a unit of time. In our simulations 1 MCS corresponds to 1 hour of real time and the parameters in the multiscale model are scaled accordingly

\begin{figure}[t!]
  \begin{center}
      \includegraphics[scale=.5]{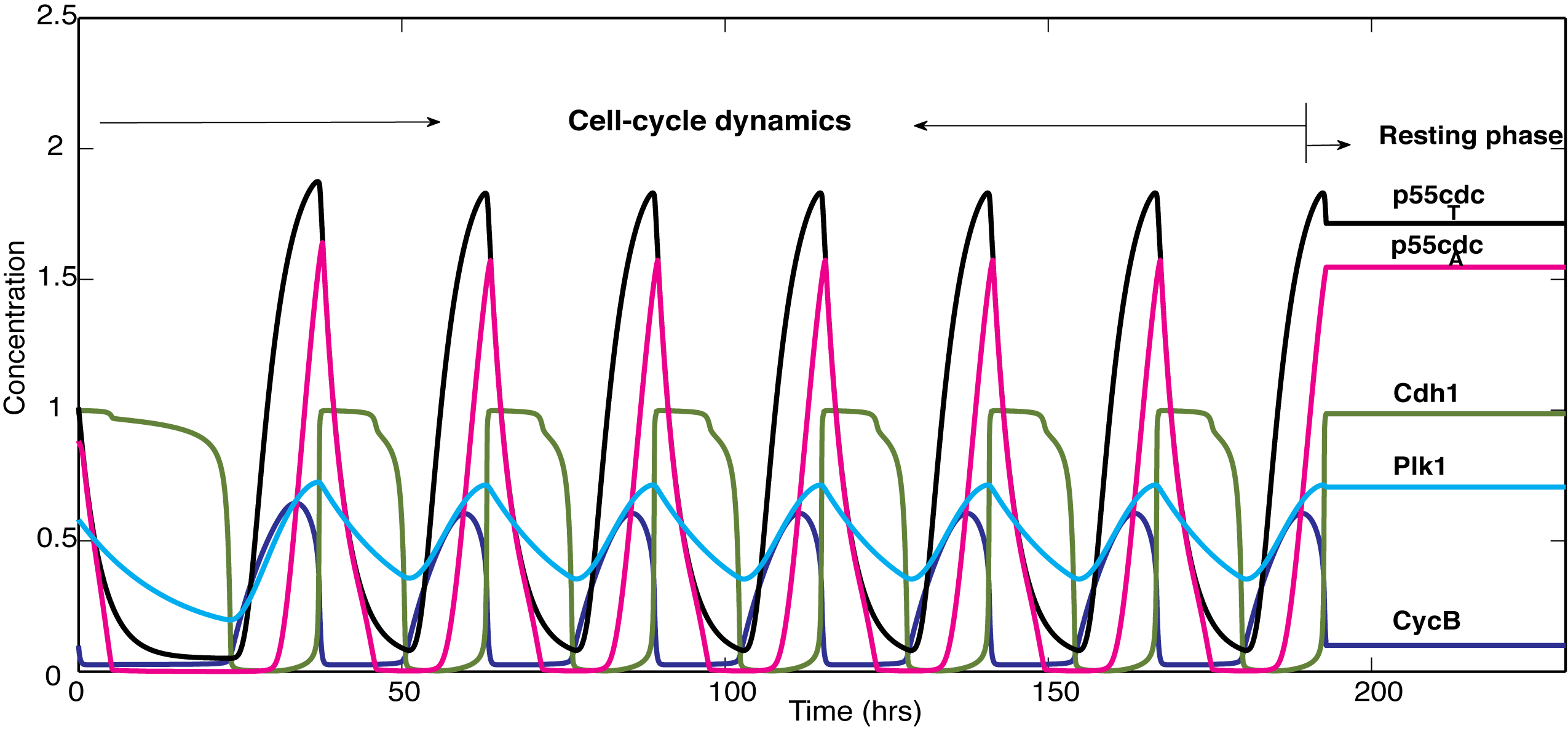}
      \end{center}
  \caption{A representative diagram showing the cell-cycle dynamics within each cell. Plots show the concentration profiles of the various intracellular proteins modelled in Equations 1-5 over a period of 200 hours.}
  \label{cellcycle}
\end{figure}
%%%

\subsection*{Cell-cycle dynamics and intracellular heterogeneity}
The growth and division of cancer cells in this hybrid multiscale model is modelled using a system of ODEs that describes the changes in the concentrations of some of the key proteins that are involved in the cell-cycle mechanism \cite{Novak2001, Novak2003}. Here, the cell-cycle dynamics within each cell are incorporated into the multiscale framework using a minimal model involving 5 variables, originally developed by Tyson and Novak \cite{Novak2001, Novak2003} that includes only the interactions which are considered to be essential for cell-cycle regulation and control. Using the kinetic relations they explained the transitions between the two main states, G1 and S-G2-M, of the cell-cycle, which is (in their model) controlled by changes in cell mass. For further details and analysis of the model, readers are referred to the papers by Tyson and co-workers \cite{Novak2001, Novak2003}. Here, we have used the equivalent mammalian proteins stated in Tyson and Novak's paper to make the current six variable model more relevant to the mammalian cell, the Cdk-cyclin B complex [CycB], the APC-Cdh1 complex [Cdh1], the active form of the p55cdc-APC complex [p55cd$\text{c}_\text{A}$], the total p55cdc-APC complex [p55cd$\text{c}_\text{T}$], the active form of the Plk1 protein, [Plk1] \cite{Powathil2012b}. The transition between phases is constrained by the cell's area, expressed in the number of pixels (or volume in 3D) that increases after each MCS. The system of ODEs governing the cell-cycle dynamics is given by:

%%%%%
\begin{align}
&\frac{d\text{[CycB]}}{dt}=k_1-(k^{'}_2+k^{''}_2\text{[Cdh1]}+[\text{p27/p21}][HIF])\text{[CycB]},\\
&\frac{d\text{[Cdh1]}}{dt}=\frac{(k^{'}_3+k^{''}_3\text{ [p55cdc}_\text{A}])(1-\text{[Cdh1]})}{J_3+1-\text{[Cdh1]}}-\frac{k_4\text{[area][CycB][Cdh1]}}{J_4+\text{[Cdh1]}},\\
&\frac{d \text{ [p55cdc}_\text{T}]}{dt}=k^{'}_5+k^{''}_5\frac{(\text{[CycB][area]})^n}{J^n_5+(\text{[CycB][area]})^n}-k_6 \text{ [p55cdc}_\text{T}],\\
&\frac{d \text{ [p55cdc}_\text{A}]}{dt}=\frac{k_7\text{[Plk1]}(\text{ [p55cdc}_\text{T}]-\text{ [p55cdc}_\text{A}])}{J_7+\text{ [p55cdc}_\text{T}]-\text{ [p55cdc}_\text{A}]}-\frac{k_8[Mad]\text{ [p55cdc}_\text{A}]}{J_8+\text{ [p55cdc}_\text{A}]}-k_6\text{ [p55cdc}_\text{A}],\\
&\frac{d\text{[Plk1]}}{dt}=k_9\text{[area][CycB](1-[Plk1])}-k_{10}\text{[Plk1]},
\end{align}

\noindent where $k_i$ are the rate constants, their values being chosen in proportion to those in Tyson and Novak \cite{Novak2001, Novak2003} as given in Powathl et al. \cite{Powathil2012b}. Other parameters used in the system are $J_i$, [Mad] and [p27/p21]. In two-dimensional CC3D on Cartesian lattice simulations the total number of pixels associated with each cell (expressed in CC3D Python code as {\it cell.volume}) is numerically equal to the cell's area. Hence in our CC3D code we use the total number of pixels to refer to the area. Similarly, in two-dimensional CC3D simulations on a Cartesian lattice, the CC3D Python statement {\it cell.surface} refers to a cell's perimeter and is numerically equal to the number of interfaces between a given cell and its neighbours -- including the medium. The effects of changes in oxygen dynamics are included through the activation and inactivation of HIF-1 $\alpha$ which further results in changes in cell-cycle length. Here, we have assumed that  HIF-1 $\alpha$ concentration at position $\Omega$, which is normally inactive ([HIF] = 0), is activated ([HIF] = 1) if the oxygen concentration at that position falls below 10$\%$. The cell-cycle inhibitory effect of p21 or p27 genes expressed through the activation of HIF-1 $\alpha$ is incorporated into the equation governing our generic Cyclin-CDK dynamics, using an additional decay term proportional to the concentration of p27/p21 (which is considered here as constant) \cite{Alarcon2004}. A representative figure illustrating the details of the cell-cycle dynamics is shown in Figure \ref{cellcycle}.

For simulations involving tumour growth, the cell's target area (expressed in CC3D Python as {\it cell.targetVolume}) is incremented in each MCS during the growth phases by 0.5 unit to achieve the assumed average target area within the individual cell-cycle time. The parameter values of the cell-cycle model are scaled in such a way that each MCS step corresponds to 1 hour in real time and a cancer cell has an average cell-cycle length of 25-35 hours. Moreover, as described in the above section, the cell-cycle model is incorporated into the CompuCell3D framework at each MCS using Bionetsolver and the SMBL ODE solver library. Note that, although here we have used a simplistic model for cell-cycle dynamics, one could easily replace this with a more complicated system incorporating more details of all the proteins and signalling molecules involved in the control of the cell-cycle. 
%%%%%%%

\subsection*{The tumour microenvironment}

Overall tumour progression and cell-cycle dynamics are critically dependent on the surrounding tissue microenvironment and in particular, the availability of oxygen. Hypoxia, a state where the local oxygen concentration falls below a critical level (i.e. 5-10 mm Hg) may result in the up-regulation of some of the intracellular pathways such as p21 or p27 pathways, which may in turn interact with the proteins involved in the cell-cycle mechanism, affecting its temporal and spatial progression \cite{Goda2003b, Gardner2001}. Here, the temporal and spatial evolution of the oxygen concentration distribution is modelled using the following partial differential equation \cite{Powathil2012b} and is incorporated into the CompuCell3D framework as a diffusive chemical field:

\begin{align}
&\frac{\partial K(x,t) }{\partial t}=\nabla.(D_K(x)\nabla K(x,t))+r(x) m(\Omega)-\phi K(x,t) \text{cell}(\Omega,t)
\label{oxygen_equation}
\end{align}

\noindent where $K(x,t)$ denotes the oxygen concentration at (pixel) position $x$ at time $t$, $D_K(x)$ is the diffusion coefficient and $\phi$ is the average rate of oxygen consumption by a cell with an area (in 2 dimensions) $\Omega$ at time $t$ ($\text{cell}(\Omega,t) =1$ if position $x\in \Omega$ is occupied by a cancer cell at time $t$ and zero otherwise). Here, $m(\Omega)$ denotes the vessel cross section with an area $\Omega$ ($m(\Omega) =1$ for the presence of blood vessel at position $x \in \Omega$, and zero otherwise); thus the term $r(x)  m (\Omega)$ describes the production of oxygen at an average rate $r(x)$ (averaged over the vessel cross section area). This equation is solved using no-flux boundary conditions and an appropriate initial condition \cite{Powathil2012}. We used an explicit Forward Euler numerical integration method to solve the PDEs governing the oxygen and drug dynamics. To avoid numerical instabilities and ensure consistency with previous work (Powathil et al. \cite{Powathil2012b}) the oxygen dynamics are simulated 1000 times in each MCS and used the same parameter values as in \cite{Powathil2012b}. We assume that hypoxia up-regulates the HIF-1 (Hypoxia Inducible Factors) pathway, which further results in changes in intracellular cell-cycle dynamics. When the average oxygen concentration at a specific cell location $\Omega$ falls below 10$\%$ (hypoxic cell), HIF-1 $\alpha$  is assumed to become active from an inactive phase, which further delays the cell-cycle dynamics (cf. Equation (1)).
%%%%%%%

\subsection*{Cell-cycle phase-specific chemotherapy}
Chemotherapy is one of the most common therapeutic options for cancer treatment, either alone or in combination with other therapies. Most of the currently used chemotherapeutic drugs act on rapidly proliferating cells by targeting the different cell-cycle phases and check points and hence, are cell-cycle specific. In cancer, the proteins responsible for the activation of the cell-cycle, Cdks, are over-expressed, while cell-cycle inhibitory proteins are under-expressed, leading to uncontrollable growth. The rationale behind cell-cycle specific chemotherapeutic drugs is to target those proteins that are over-expressed during various stages of cancer progression, inducing an inhibitory effect by blocking the cell division cycle at a specific phase. Consequently, cell-cycle-specific chemotherapeutic drugs are more effective on dividing cells by interfering with the cell-cycle and other cell-cycle specific targets. However, while the concept of phase-specific chemotherapy is useful, and although some drugs have specific effects on the machinery of mitosis, it is becoming clear that chemotherapy drugs may affect more than one aspect of the cell cycle, and so the concept of phase-specificity is somewhat of an over-simplification. Here, we are interested in investigating the effects of cell-cycle specific chemotherapy on multiple cancer cell populations with varying drug sensitivity. The spatio-temporal distribution of the chemotherapeutic drugs is governed by a reaction-diffusion equation similar to that of the oxygen concentration, given by:

\begin{align}
&\frac{\partial C_i(x,t) }{\partial t}=\nabla.(D_{ci}(x)\nabla C_i(x,t))+r_{ci}(x) m(\Omega)-\phi_{ci} C_i(x,t) \text{cell}(\Omega,t)-\eta_{ci}C_i(x,t)
\label{chemo_equation}
\end{align}

\noindent where $C_i(x,t)$ is the concentration of chemotherapeutic drug type $i$, $D_{ci}(x)$ is the diffusion coefficient of the drug,  $\phi_{ci}$ is the rate by which the drug is used up by a cell (assumed to be zero as it is very negligible when compared to oxygen uptake),  $r_{ci}$  is the drug supply rate by the pre-existing vascular network and $\eta_{ci}$ is the drug decay rate \cite{Powathil2012b}. Here, we assume a uniform drug supply throughout the domain, regardless of the spatial location of the blood vessel. To study the regrowth after chemotherapeutic treatment we have assumed that the given drug does not eradicate the tumour completely. Moreover, we have used same threshold of effectiveness for the given chemotherapeutic drugs to compare the results appropriately.

\section*{Results and Discussion}

Although a number of new anti-cancer drugs that target various key processes in cancer progression have been under development, overall patient survival rates have mostly remained unchanged for a number of years. One of the major factors that contributes to this is evolved drug resistance and this in turn is due in part to the presence of multiple subpopulations of cancer cells with varying cell-cycle dynamics and cell-cycle length that evolve within a heterogenous tumour microenvironment \cite{Bailar1997}. Clinical and experimental observations have indicated that the correct timing and dosing of chemotherapeutic drugs as well as their optimal sequencing may help to reduce the tumour burden and further help to increase the survival status of patients \cite{Masson1997}. Even with the presence of heterogenous subpopulations of cancer cells with varying cell-cycle dynamics, a better understanding of the cell-cycle mechanism may help in the optimisation of multi-drug chemotherapy delivery involving cell-cycle phase-specific chemotherapeutic drugs \cite{Deep2008, Zalatnai}. In this section, we study the growth and evolution of a heterogenous tumour mass that consists of subpopulations of cancer cells with varying cell-cycle lengths in order to understand the role of intracellular perturbations in the cell-cycle dynamics in inducing chemotherapeutic drug resistance. We also discuss the effectiveness of cell-cycle phase-specific chemotherapeutic drugs in controlling the progression of the heterogeneous tumour mass. 

All the simulations begin with a single cancer cell in the G1 phase (a blue cell) of the cell-cycle at the centre of the domain. This initial cell divides repeatedly following the ODEs governing the cell-cycle dynamics and eventually produces a cluster of cancer cells. Following the Cellular Potts Model methodology \cite{Glazier2007} we measure time in units of Monte Carlo Steps (MCS). In our model a single MCS corresponds to 1 hour of real time. At each MCS, all the cells are checked for the concentrations of intracellular protein levels and their phases are updated. If [CycB] is greater than a specific threshold (i.e. $0.1$) the cell is considered to be in G2 phase (a green cell) and if it is lower than this value, the cell is in G1 phase. If the [CycB] crosses this threshold from above, the cell undergoes cell division and divides along randomly chosen cleavage plane. As the cells proliferate, the oxygen demand increases, and in some regions the concentration of oxygen falls below a certain threshold value (10\% of oxygen), making the cells hypoxic and hypoxic cells are further assumed to have a longer cell-cycle due to the cell-cycle inhibitory effect of p21 or p27 genes expressed through the activation of HIF-1 under hypoxia. Moreover, if the oxygen level of a cancer cell falls below 1\%, the cell is assumed to enter into a resting phase with no active cell-cycle dynamics \cite{Goda2003, Goda2003b}. The computational  domain of size $300\times300$ will be able to simulate a cancerous mass of area $1\times1$ mm$^2$, approximately. Although the size of the tumour simulated here does not always reflect the size of {\it in vivo} tumours that are seen in clinical situations (which may be larger), qualitative insights can still be gained from the results of the computational simulations. Furthermore, such hybrid mathematical models at this spatial scale can be very beneficial in helping to understand and study the various nonlinear factors affecting chemotherapeutic drug resistance in tumours.

\subsection*{Cancer progression with homogeneous cell-cycle dynamics}

\begin{figure}[t!]
  \begin{center}
      \includegraphics[scale=1.15]{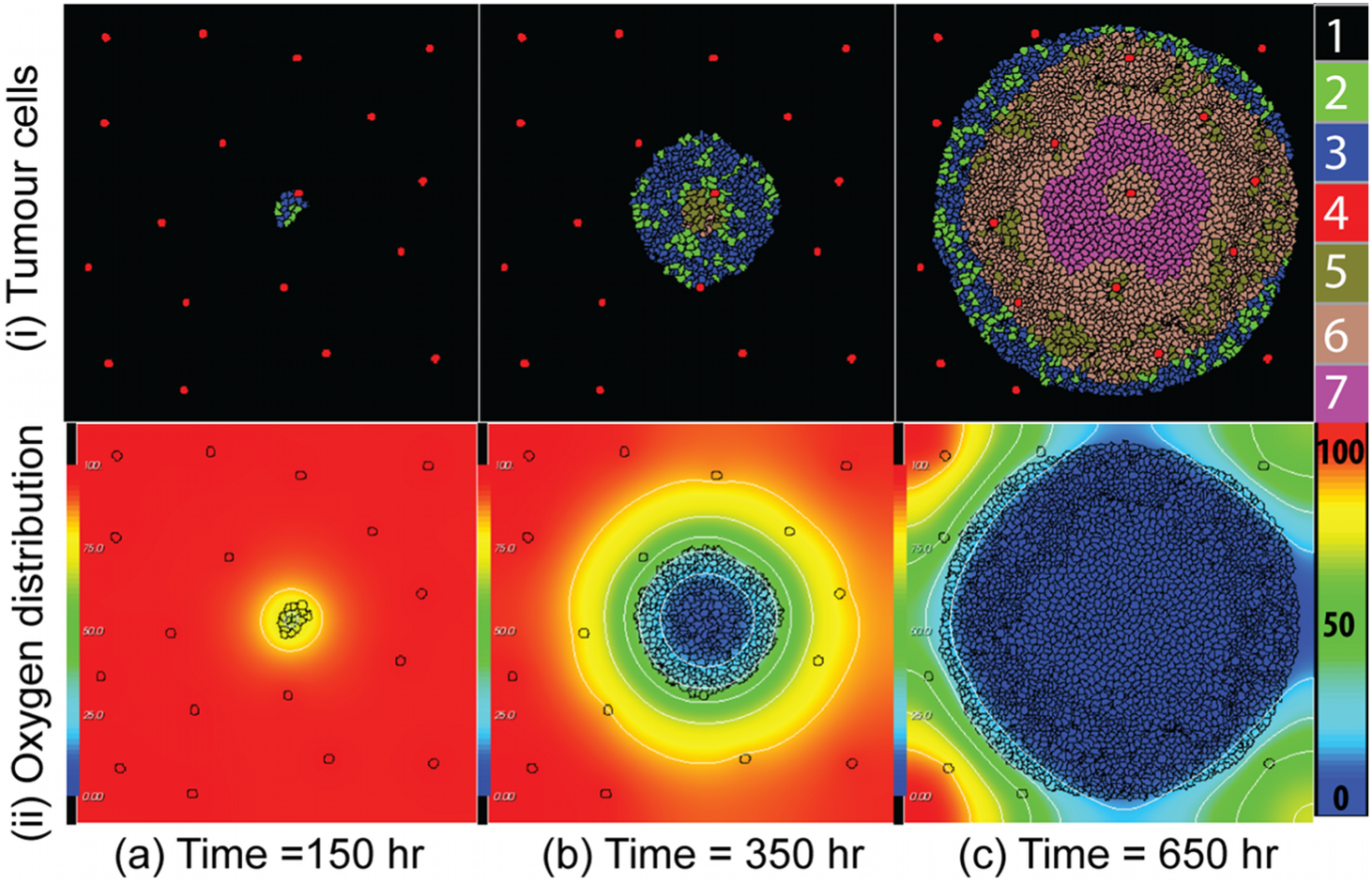}
      \end{center}
  \caption{Plots showing snapshots of the computational simulation results of the model at various time points. (i) Plots of the spatial evolution of cancer cells in different phases of the cell-cycle at times  150 hrs, 350 hrs and 650 hrs. The colour legend shows the different types of cancer cells and cells of the microenvironment; 1 - medium, 2 - S-G2-M phase, 3 - G1 phase, 4 - blood vessel cross sections, 5 - hypoxic S-G2-M phase, 6 - hypoxic G1 phase and 7 - resting cells. (ii) Plots of spatial evolution of oxygen concentration at times 150 hrs, 350 hrs and 650 hrs.}
  \label{cells}
\end{figure}

\begin{figure}[t!]
  \begin{center}
      \includegraphics[scale=1.15]{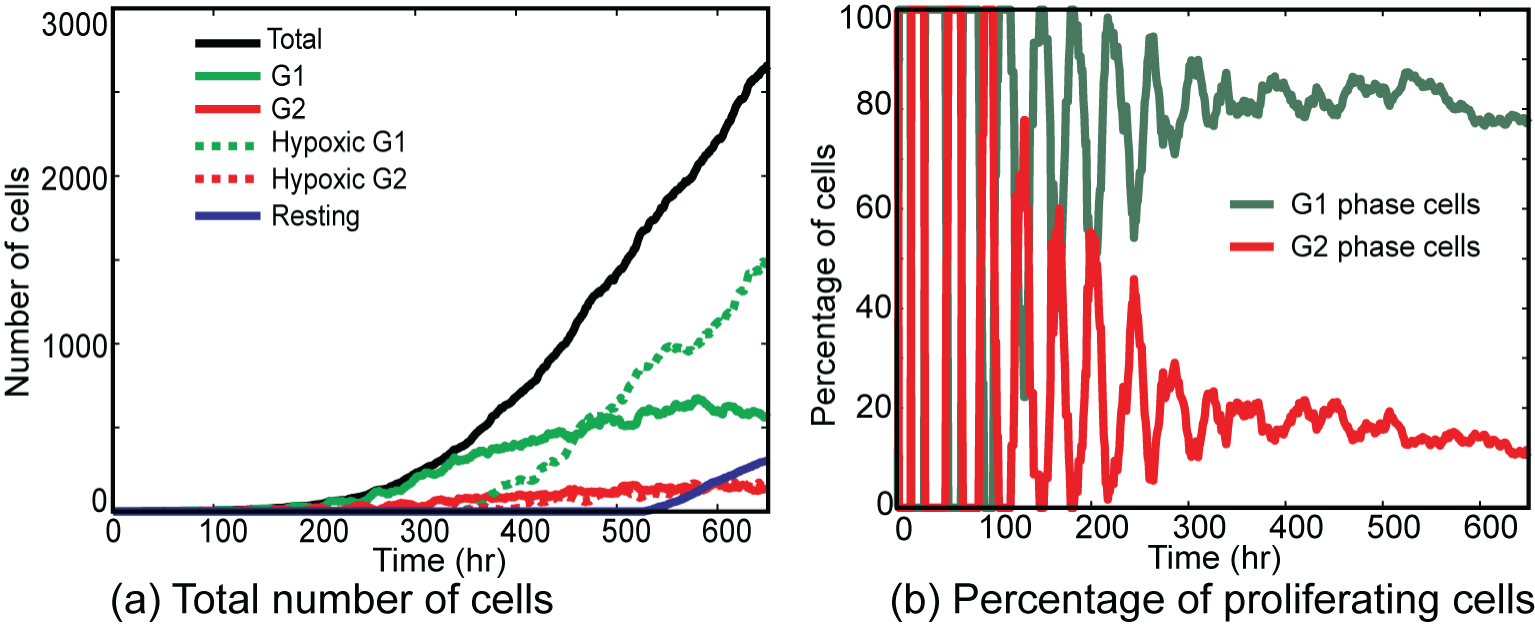}
      \end{center}
  \caption{Plots showing (a) the number of cancer cells in various phases of the cell-cycle over a period of 650 hrs and (b) the percentage of proliferating cancer cells either in G1 phase or S-G2-M phase of the cell-cycle over a time of 650 hrs.}
  \label{cells_no}
\end{figure}

In this section we discuss the computational simulation results of our hybrid multiscale model for a homogenous cancer cell population but incorporating intracellular and microenvironmental variations. Figure \ref{cells} shows the spatial distribution of the cancer cells and oxygen concentration at three different time points. The spatial evolution of cancer cells is given in Figures \ref{cells} (i: a-c) and the spatial distribution of the oxygen concentration is given in Figures \ref{cells}(ii: a-c). The colour of the cancer cells indicates the specific phase position they occupy in the cell-cycle. Oxygen concentration and microenvironment status (e.g. external medium, blood vessels) is also shown by different colours (see figure legend for details). From these plots it can be seen that initially the tumour mass is composed only of proliferating cancer cells with active cell-cycle dynamics that cycle between G1 phase and S-G2-M phase. As time increases, the oxygen consumption by the cancer cells increases and the actively proliferating cells are seen only at the tumour boundary while hypoxia starts developing within the tumour. The total number of cancer cells and the number of cancer cells that are in G1, S-G2-M, and the resting phases are given in Figure \ref{cells_no} (a), while the percentages of proliferating cancer cells in G1 phase and S-G2-M phase are shown in Figure \ref{cells_no} (b). In the early stages of progression, the cells in G1 phase and S-G2-M phase cycle synchronously and as time increases the percentage of cells in G1 phase dominates those in S-G2-M phase and tends to stay around a steady-state level as seen in several experimental studies \cite{Chaudhry2007, goto2002}. From Figure \ref{cells_no} (a), it can be seen that with the presence of hypoxia, the number of cancer cells in the hypoxic G1 phase increases due to the expression of the cyclin-dependent kinase-inhibitors p21 and p27 that are up-regulated under hypoxic conditions, resulting in a prolonged cell-cycle time or even cell-cycle arrest \cite{Goda2003b, Gardner2001}. 

%%%%
%%%%

 %%%%
%%%

\subsubsection*{Effects of cell-cycle phase-specific chemotherapeutic drugs}

\begin{figure}[t!]
  \begin{center}
      \includegraphics[scale=1.25]{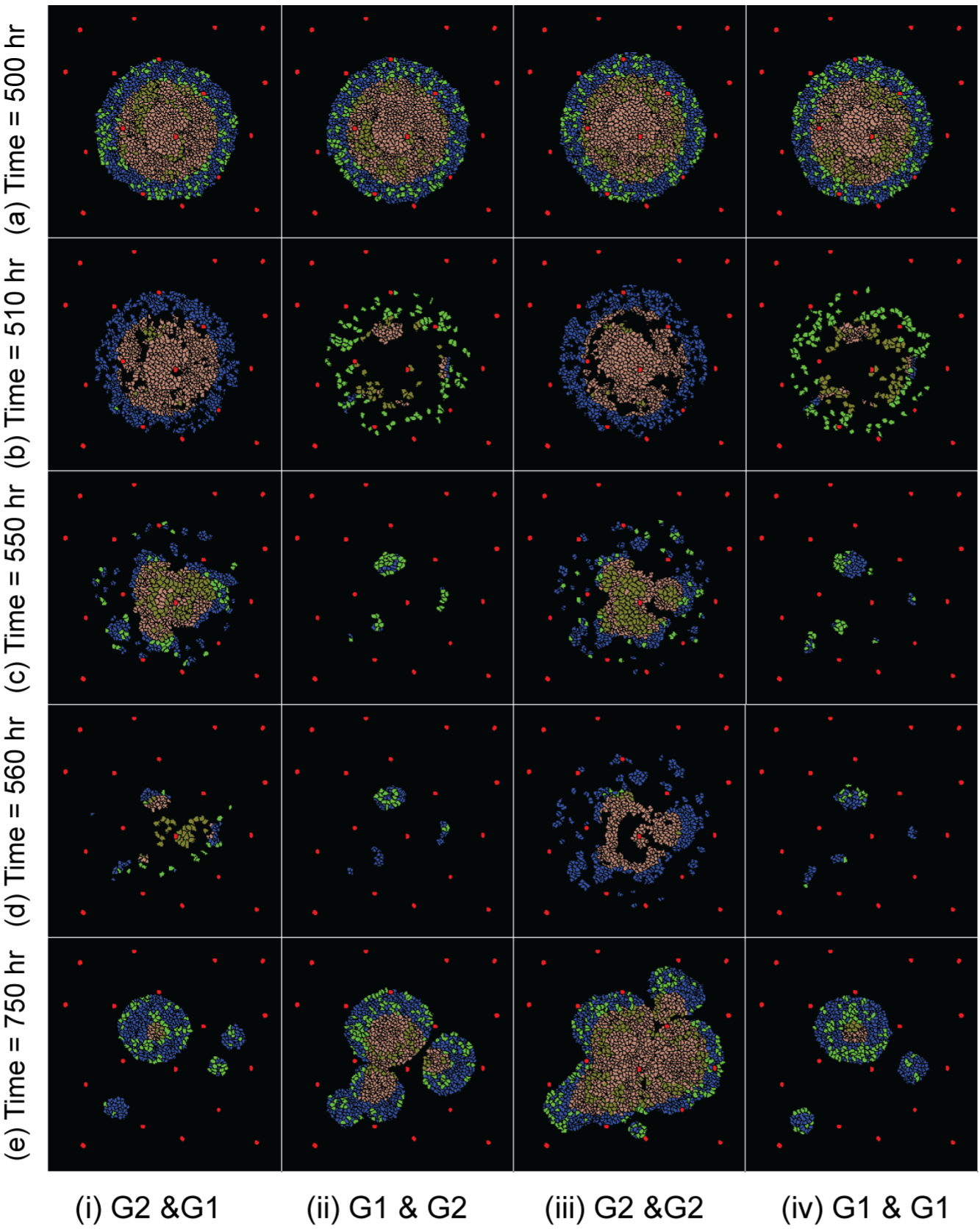}
      \end{center}
  \caption{Plots showing the spatial evolution of cancer cells when various combinations of cell-cycle phase-specific chemotherapeutic drugs are given. (i) S-G2-M drug followed by G1 drug, (ii) G1 drug followed by S-G2-M drug, (iii) S-G2-M drug followed by S-G2-M drug and (iv) G1 drug followed by G1 drug. Times (a) 500 hr, (b) 510 hr, (c) 550 hr, (d) 560 hr and (e) 750 hr.}
  \label{homo_treat}
\end{figure}

\begin{figure}[t!]
  \begin{center}
      \includegraphics[scale=1]{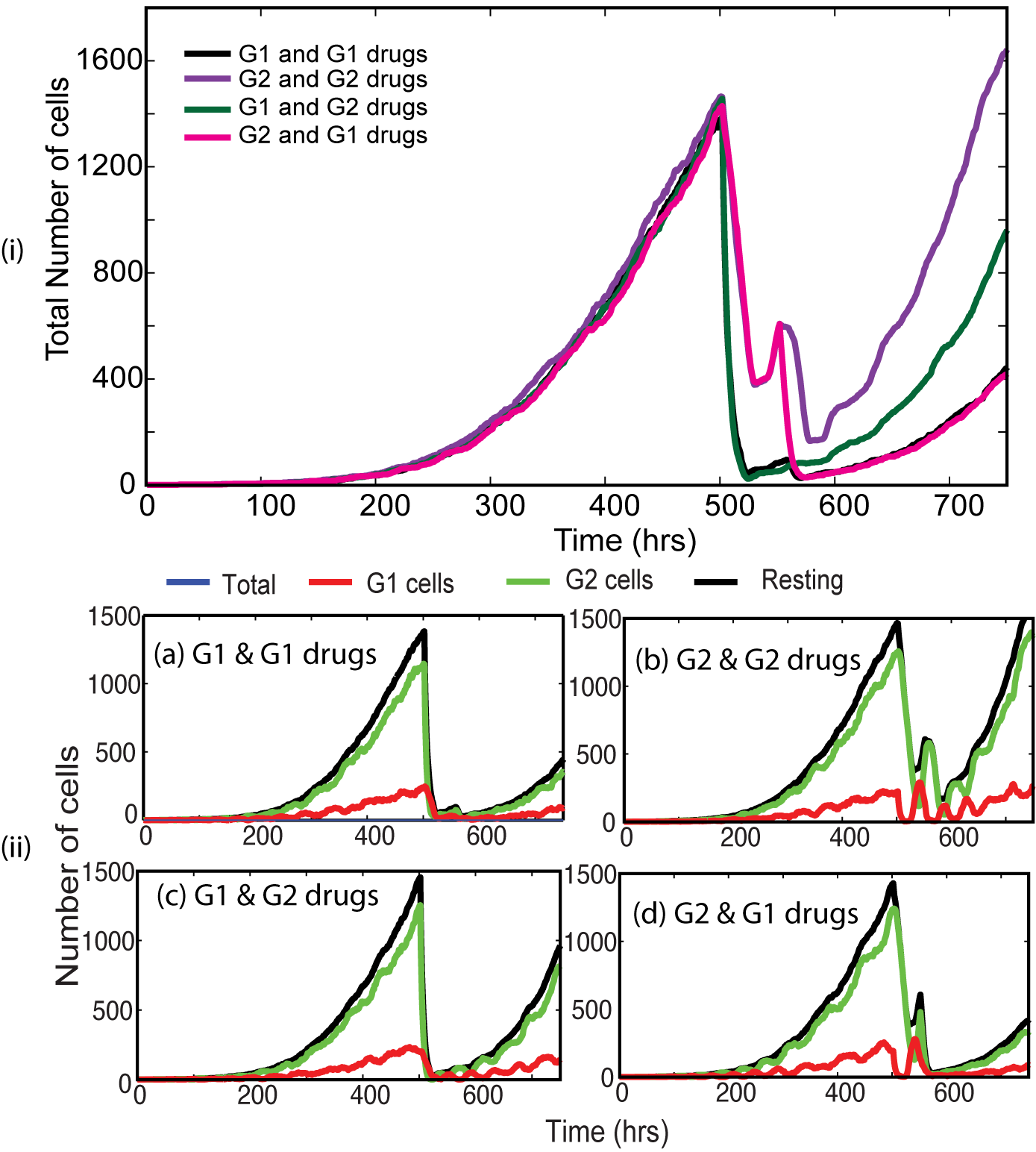}
      \end{center}
  \caption{Plots comparing the number of cells when the tumour cells are treated with two doses of cell-cycle phase-specific drugs at Time = 500 hr and Time = 550 hr. (i) Plot showing the total number of cells and (ii) Plots showing the total number of cells, cells in G1 phase, S-G2-M phase and resting phase for four different combinations of phase-specific chemotherapeutic drugs.}
  \label{homo_treat_no}
\end{figure}

Most chemotherapeutic drugs act on rapidly proliferating cancer cells that are either in the G1 phase or S-G2-M phase of the cell-cycle. Consequently, an optimal use of phase-specific chemotherapeutic drugs can be used effectively to increase the cytotoxic effects on chemotherapeutic cells. Previously, Powathil et al. \cite{Powathil2012b} showed that the spatial distribution of cancer cells in the tumour cell mass, the timing of the drug delivery, the time between the doses of cytotoxic drugs, and the cell-cycle and oxygen heterogeneity play important roles in determining the precise cytotoxic effectiveness of cell-cycle phase-specific chemotherapeutic drugs. In order to understand the cell-cycle redistribution patterns after delivery of the chemotherapy, two doses of cell-cycle phase-specific drugs that act on cells that are either in G1 phase (e.g. CYC202 or Cisplatin) or S-G2-M phase (e.g. Taxotere or Taxol) are delivered at a same rate at times 500 hours and 550 hours. Figure \ref{homo_treat} shows the temporal and spatial evolution of the cancer cells when the tumour is treated with the chemotherapeutic drugs. The comparison between the total number of cancer cells and cells in various phases of the cell-cycle is given in Figure \ref{homo_treat_no}.

 %%%%
%%%
 %%%%
%%%

Figures \ref{homo_treat}(a)-(e) shows the spatial distribution of cancer cells during the administration of various combinations of phase-specific drugs at times 500 hrs (before administration of the phase-specific chemotherapeutic drugs), 510 hrs (10 hrs after the delivery of dose 1), 550 hrs, 560 hrs (10 hrs after the delivery of dose 2) and 750 hrs (the final time point of the simulation) respectively. The plots in column (i) show the combination of a S-G2-M phase and a G1 phase-specific drug, column (ii) the combination of G1 phase and S-G2-M phase drugs, column (iii) the combination of S-G2-M phase and S-G2-M phase drugs and finally column (iv) the combination of G1 phase and G1 phase drugs. As can be seen from the results, the spatial distribution of cancer cells and the internal spatial structure of the overall tumour mass varies greatly depending on which specific combination of phase-specific drugs is given. This result in itself shows the importance of an explicitly spatial multi-scale approach. 

The results of figure \ref{homo_treat} concerning the different spatial distributions of the cancer cells over time clearly has implications for the total number of cancer cells present at any given time and also for the number in any given phase of the cell-cycle. Figure \ref{homo_treat_no} (i) shows the comparison between the total number of cancer cells over time in each of the previous combination scenarios. The plots show that after a time of 650 hrs, the two combinations when a G1 phase-specific drug is given as the second dose give a better result than other two combinations. The changes in the number of cells in various cell-cycle phases for the four combinations of chemotherapeutic delivery are given in Figure \ref{homo_treat_no} (ii). Since a higher fraction of proliferating cells are in G1 phase, the delivery of G1 phase drug kills a higher number of cancer cells. After killing the cancer cells, the reduced tumour burden changes the tumour microenvironment which in turn increases tumour oxygenation and reduces hypoxia, thus finally increasing cancer cell proliferation. Therefore, when the next dose of the phase-specific drugs is delivered at time = 550 hrs, most of the cells are in G1 phase and hence a dose of G1 phase drug will be more effective than the S-G2-M phase drug (Figure \ref{homo_treat_no} (ii)). In the presented scenario, most of the cancer cells are proliferating unless there is severe hypoxia (less than 1\% of oxygen), and so combinations of S-G2-M phase \& G1 phase drugs and G1 phase \& G1 phase drugs give a better outcome than other two combinations. However, previous studies have shown that \cite{Powathil2012b}, these treatment regimes need not necessarily give the best outcome, especially if there were a higher proportion of resting cells within a growing tumour mass. Once again this shows the importance of knowing the precise spatial structure of a growing tumour mass and the internal cellular heterogeneities present. This also indicates that the choice of an optimum therapeutic regime clearly depends on the proportion of cells in various phases of the cell-cycle, showing the importance of multiscale modelling in patient-specific treatment protocols. 

\subsection*{A heterogeneous tumour mass with a slow-cycling subpopulation of cells}

Drug resistance to chemotherapy is one of the major hurdles to the success of various cancer treatment protocols \cite{Shah2001}. Although some of the reasons for the emergence of drug resistance within a growing tumour are already known, such as over-expression of drug transporters, changes in drug kinetics and amplification of drug targets, other reasons for the drug resistance are largely unknown \cite{Saunders2012, Koshkin2012}, and unfortunately tumour recurrence after chemotherapy is a known reality. Recently, intra-tumoural heterogeneity including active intra-cellular cell-cycle dynamics has been shown to contribute towards the development of cancer cell resistance to various chemotherapeutic drugs. Additionally, most solid tumours are comprised of multiple variants of tumour subpopulations as cancer cells evolve over time with continuous genetic mutations that are actively passed on to daughter cells as they divide, resulting in a multi-clonal, heterogeneous solid tumour mass with varying characteristics \cite{Saunders2012}. It has been observed that most of the traditional chemotherapeutic drugs, such as 5-FU and Oxaliplatin, are much more effective against actively cycling cells, with slow-cycling or quiescent cells being less susceptible to these drugs \cite{Moore2012, Moore2011, Moore2012b,Schillert2012}. This leads to an emerging subpopulation of slow-cycling cells that has the ability to repopulate the tumour with potentially drug-resistant cells. Recently, several types of cancer have been identified as containing subpopulations of slow-cycling cells. Some specific examples include the PKH26 label-retaining subpopulation that has a doubling time of 4 weeks and is found in primary melanoma cell lines \cite{Roesch2010}, slow-cycling subpopulations in ovarian cancer \cite{Kusumbe2009} and slow-cycling subpopulations found in pancreatic adenocarcinoma \cite{Dembinski2009}. The slow-cycling feature of these cancer cells is sometimes associated with the characteristics of stem-like cells and there is also evidence that normal and neoplastic cells can sometimes de-differentiate into stem cells with slow-cycling properties \cite{Moore2012, Moore2011, Chaffer2011}. 

%%%%
\begin{figure}[t!]
  \begin{center}
      \includegraphics[scale=1.4]{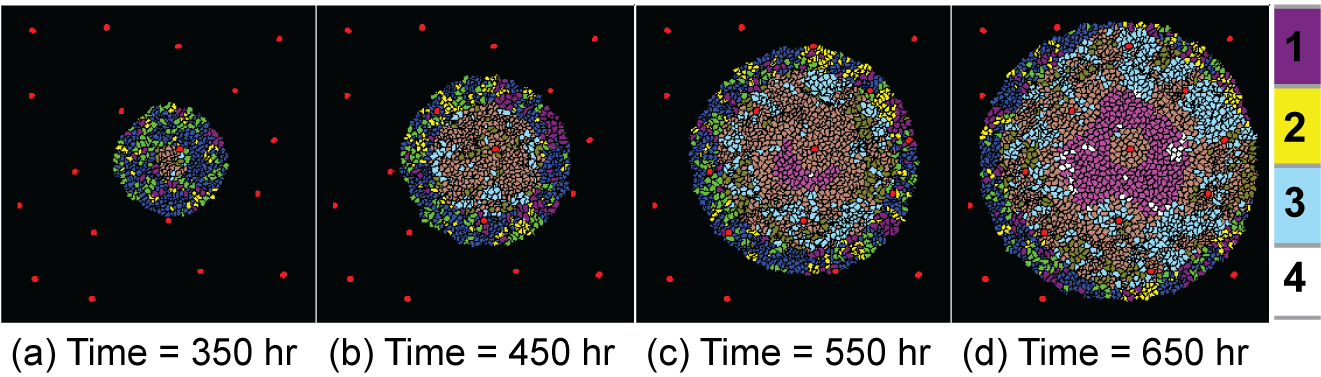}
      \end{center}
  \caption{Plots showing snapshots of the simulation results of the model with two subpopulations of cancer cells at times (a) 350 hrs, (b) 450 hrs, (c) 550 hrs and (d) 650 hrs. The colour legend given in addition to that of Figure \ref{cells} (i) (for subpopulation 1) shows the specific phases of the cell-cycle and types of cancer cells in the second subpopulation: 1 - S-G2-M phase, 2 - G1 phase, 3 - hypoxic and 4 - resting cells.}
  \label{cells2}
\end{figure}
%%%

\subsubsection*{Heterogenous model: Comparison with experimental results}
%%%%
\begin{figure}[t!]
  \begin{center}
      \includegraphics[scale=0.85]{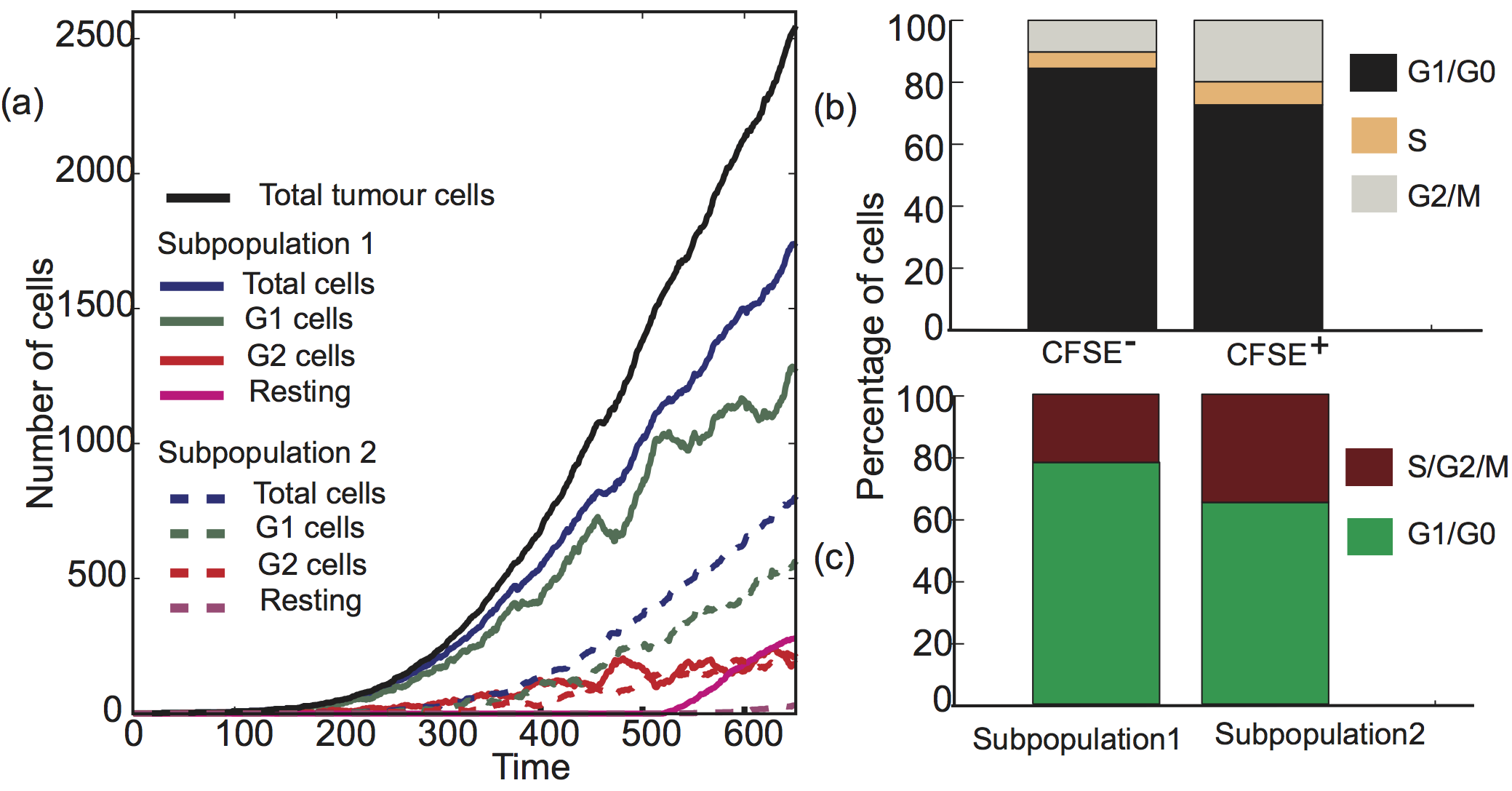}
      \end{center}
  \caption{Plots showing (a) the number of cancer cells in various phases of the cell-cycle for a heterogenous tumour mass with a second subpopulation of slow-cycling cancer cells; (b) experimental data (taken from Moore et al. \cite{Moore2012}) showing the percentage of cells in either G1 phase or S-G2-M phase for CFSE $^-$ and CFSE$^+$subpopulations; (c) simulation results showing the average percentage of cells in G1/G0 and S/G2/M phases for subpopulations 1 and 2.}
  \label{cells2_no}
\end{figure}
%%%
To study the effect of chemotherapeutic drugs on a growing tumour mass that contains slow dividing cancer cells, we consider a two-subpopulation model with one subpopulation taking twice as much time as the other to complete one cell-cycle under favourable conditions. Initially, the tumour is assumed to evolve with similar dynamics as that of the single population model explained in the previous section, with a cell-cycle length of between 25-30 hrs for all cells for the first 100 MCS. Mutations are introduced into this growing tumour after 100 MCS with certain probabilities to form a slow dividing subpopulation of cancer cells that grows alongside the other (faster dividing) cancer cells. Previously, Goldie and Coldman \cite{Goldie1979, Goldie1982, Goldie1983, Goldie1985, Coldman1985} proposed that cancer cells mutate to a drug-resistant phenotype at a rate dependent on their intrinsic genetic instability. They showed that even the smallest fraction of a solid tumour contains at least one drug-resistant subpopulation. Here, we assume that mutations giving rise to a slow-cycling subpopulation of cells occur with a low probability of 2-3\% in every 25 hours (the average cell-cycle time). Hence the cancer cells are checked every 25 hours by generating a random number for each cell and comparing it against the assumed probability. The computational results for the two-population growth model using the multiscale approach are given in Figures \ref{cells2} and \ref{cells2_no}.

Figure \ref{cells2} shows the spatial evolution of the heterogeneous tumour population containing two subpopulation of cancer cells and Figure \ref{cells2_no} compares the number of cells in various phases of cell-cycle for the subpopulations of normal and slow-cycling cancer cells and shows the percentages of cells in the various phases of the cell-cycle for subpopulations 1 and 2. As Figure \ref{cells2} illustrates, the spatial distribution of cancer cells for the two-population model is similar to that of the single population model but with more heterogeneity. However, the simulations show that the total cell population for the heterogenous population (Figure \ref{cells2_no} (a)) is a little lower than that of the single population model, as we have assumed a mutated, slow-cycling subpopulation of cancer cells within the tumour mass. Moreover, in the fast-dividing (subpopulation 1) subpopulation of cancer cells, the number of cells in G1 phase is higher than the number of cells in S-G2-M phase as compared to the slow-dividing cells. Figures \ref{cells2_no}(c) show that, for subpopulation 2 (slow-dividing cells), the average percentage of cells in S-G2-M phase is at least 12\% higher than those in subpopulation 1, with mean values 34.9 \% (S-G2-M phase cells in subpopulation 2) compared to 22\% in subpopulation 1. These simulation results are in qualitatively very good agreement with the experiment results of Moore et al. \cite{Moore2012, Moore2012b, Moore2011}, also shown in Figure \ref{cells2_no}(b), where they used HCT116 cell lines to study the properties of slow-cycling cells and their response to chemotherapy. Using a novel proliferation marker, carboxyfluorescein diacetate (CFSE), Moore et al. \cite{Moore2012} identified a subpopulation of slow-cycling cancer cells in both {\it in vitro} sphere cultures and {\it in vivo} xenograft models. They also showed through cell-cycle analysis of the slow-cycling cells (CFSE$^{+}$) that there is a two-fold increase in G2/M phase cells as compared with the rest of the cancer cells (CFSE$^{-}$), indicating a possible longer G2/M phase or phase arrest \cite{Moore2012} (please see Figure 1 (D) and Figure 2 (D) of \cite{Moore2012} for the raw data with error bars and other details).

\subsubsection*{Effects of chemotherapeutic drugs on heterogeneous tumour: Comparison with experimental results}
%%%%%

\begin{figure}[t!]
  \begin{center}
      \includegraphics[scale=.5]{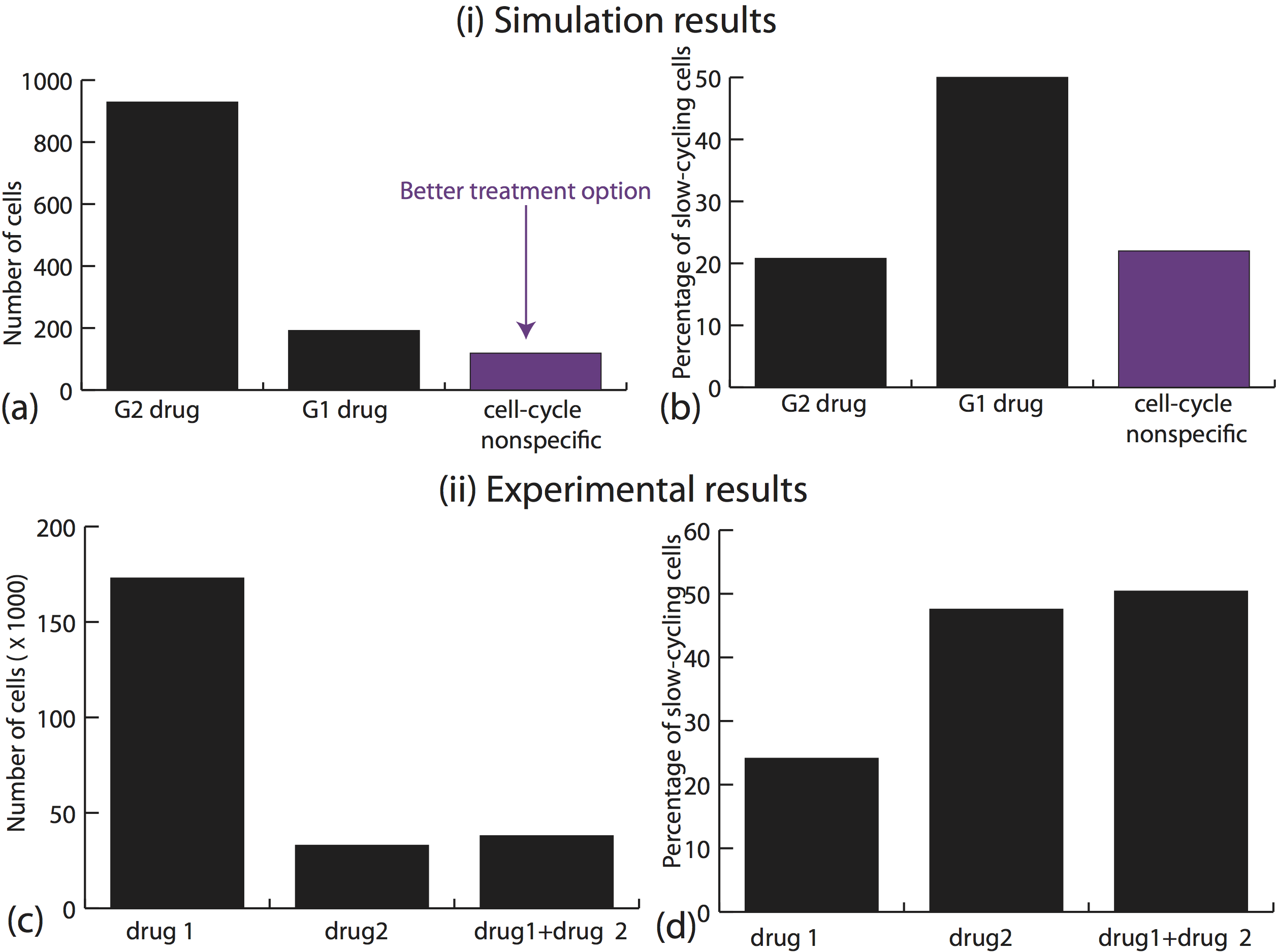}
      \end{center}
  \caption{Comparison of simulation results with the experimental results. (i) Results from the computational model at time = 580 hrs (\~ 3 days after treatment) when the cells are treated with one dose of either G1 drug, S-G2-M drug or a cell-cycle nonspecific drug that acts on all the cells at time 500 hrs. Here, (a) shows the total number of cells and (b) shows the percentage of slow-cycling cells. (ii) Experimental results from Moore et al. \cite{Moore2012} showing the (a) total number of cells and (b) percentage of CFSE$^{+}$ (slow-cycling), 3 days after the cells are treated with either drug 1 (Oxaliplatin), drug 2 (5FU) or the combination of these two with similar total dosage.}
  \label{FigureR3}
\end{figure}
%%%%	
	
%%%%%
\begin{figure}[t!]
  \begin{center}
      \includegraphics[scale=1.1]{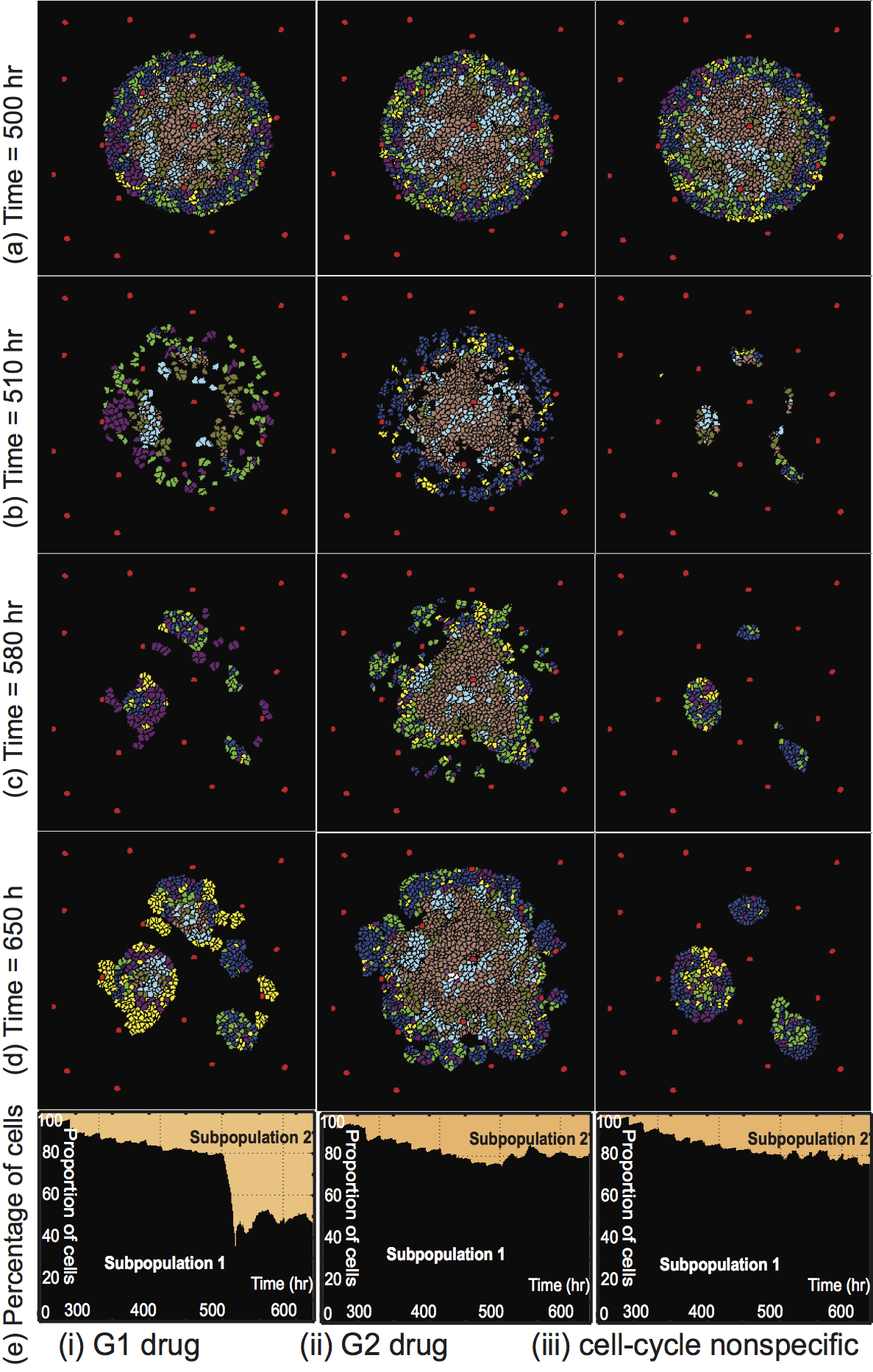}
      \end{center}
  \caption{Plots showing the spatial evolution of cancer cells within a two-population model when single dose of cell-cycle phase-specific chemotherapeutic drugs are given. (i) G1 phase drug, (ii) S-G2-M phase drug and (iii) cell-cycle nonspecific drug that acts on all the cells (a) 500 hrs, (b) 510 hrs, (c) 580 hrs, (d) 650 hrs and (e) percentage of cancer cells in subpopulations 1 and 2.}
  \label{FigureR4}
\end{figure}
%%%%

Most chemotherapeutic drugs spare some proportion of the targeted cancer cells due to their ineffectiveness in killing cells in different cell-cycle phases, including the resting phase. To study the effects of traditional chemotherapeutic drugs such as Oxaliplatin, 5-FU (5-fluorouracil) or a combination of Oxaliplatin and 5-FU, Moore et al. \cite{Moore2012, Moore2012b, Moore2011} used the HCT116 cell line which is known to include slowly-cycling cells {\it in vivo} and {\it in vitro}. Their results (original paper \cite{Moore2012}, Figure 4 (B) and (C)) show that when a tumour is allowed to grow under control conditions for 3 days, the number of cells increases 5-fold, out of which only 7.1\% of cells were slowly-cycling cells (CFSE$^{+}$ cells). However, when the cancer cells were treated with chemotherapeutic agents, the total number of viable cells remained considerably lower than the control case (Oxaliplatin - 15\% of control, 5-FU - 2.8\% of control and combination - 3.8\% of control), but the percentage of  the CFSE$^{+}$ subpopulation of cells contributing towards the total number increased (Oxaliplatin: 3.4-fold of CFSE$^{+}$ cells in control, 5-FU: 6.7-fold of CFSE$^{+}$ cells in control and combination: 7.1-fold of CFSE$^{+}$ cells in control). These results for this cell line indicate that when the tumour mass responds in a significantly favourable manner to the initial chemotherapeutic treatment, most of the chemo-responsive cancer cells are killed, increasing the percentage of slowly-cycling chemo-resistant cancer cells and thus increasing the chance of subsequent treatment failure and tumour recurrence.

First, we use our multiscale computational model to study the effects of chemotherapeutic drugs acting on either the G1 phase or the S-G2-M phase of the cell-cycle, or acting on all cancer cells irrespective of the cell-cycle phase. To compare our simulation results with the experimental results of Moore et al. \cite{Moore2012, Moore2012b, Moore2011} (detailed above), a single dose of chemotherapeutic drug is given to the heterogenous tumour at time 500 hrs and its effects on the total number and the proportion of the cancer cell subpopulation on the third day is analysed.  Figure \ref{FigureR3} shows the comparison of results from the mathematical model with the experimental results of Moore et al. \cite{Moore2012, Moore2012b, Moore2011}, where the simulation results are given in Figure \ref{FigureR3}(i) and the experimental results are given in Figure \ref{FigureR3}(ii) (raw data extracted from Figure 4 of \cite{Moore2012}). Figure \ref{FigureR3}(a) shows the total number of cells when a single dose of either S-G2-M phase-specific or G1 phase-specific or a cell-cycle nonspecific drug is given and the corresponding percentage of slowly-cycling cells are plotted in Figure \ref{FigureR3}(b). Similarly, Figures \ref{FigureR3}(c) and \ref{FigureR3}(d) show the total number of cells and corresponding CFSE$^+$ (slowly-cycling) cells, when the tumour cells are given a single dose of drug 1 (Oxaliplatin) or drug 2 (5FU) or a combination of these two drugs with similar total dosage. The simulation results show that when the given drug kills an increased number of cells, the proportion of slowly-cycling cells increases, except for the case where a hypothetical cell-cycle nonspecific drug is used. This result is in qualitative agreement with the experimental results where an increased cell-kill refines the heterogenous tumour subpopulations, thereby enriching the number of slowly-cycling tumour cells and thus increasing the chemo-resistance in subsequent doses. Further simulations indicated that the extended S-G2-M phase (Figure \ref{cells2_no}) might be the possible reason for this enrichment within the heterogenous tumour, supporting the experimental inferences \cite{Moore2012,Moore2012b, Moore2011, Schillert2012}. 

The plots in Figure \ref{FigureR4}(a-d) show the spatial evolution of the cancer cells in the various phases just before the dose of either the G1 phase or S-G2-M phase drug ((a) Time = 500 hrs), after the dose ((b) Time = 510 hrs), \~3 days after the treatment ((c) Time = 580 hrs) and at a final time of simulation ((e) Time = 650 hrs). The percentage of cancer cells in subpopulations 1 and 2 for each combination of cell-cycle phase-specific drugs is given in Figure \ref{FigureR4} (f). As the Figures \ref{FigureR3} and \ref{FigureR4} indicated administering a chemotherapeutic drug that targets all the cells, irrespective of the cell-cycle phases or other intracellular heterogeneities would be the best possible option as that not only reduced the tumour mass but also targets both tumour subpopulations, preventing the enrichment of slow-cycling cells. Although such chemotherapeutic drugs are usually ideal, they are currently unavailable. The alternative option to achieve a similar outcome is to use multiple doses of chemotherapeutic drugs that target various phases of the cell-cycle and the following section explores such combination treatment scenarios.

\begin{figure}[t!]
  \begin{center}
      \includegraphics[scale=1.06]{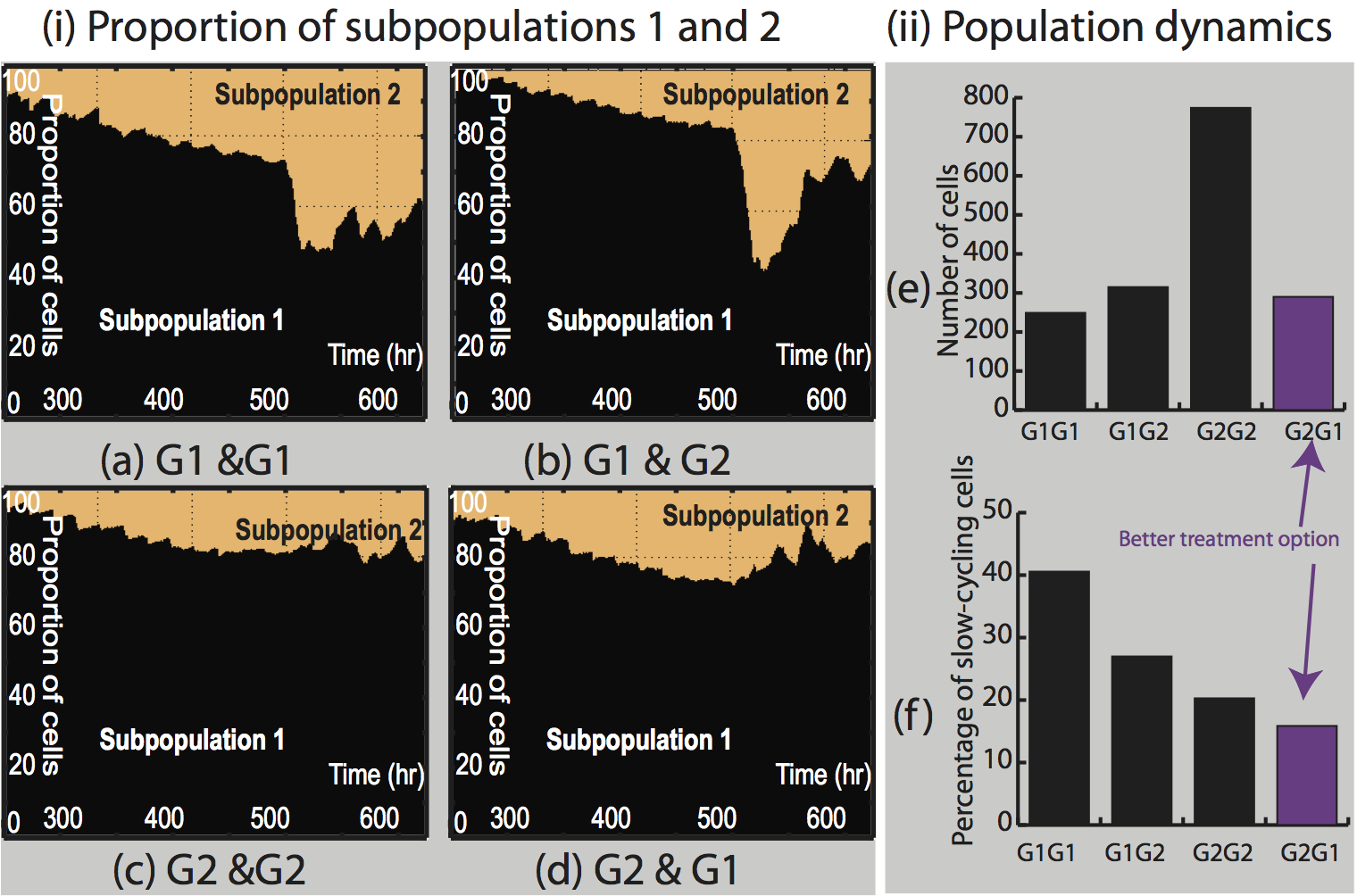}
      \end{center}
  \caption{Plots showing the (i) percentage of cancer cells in subpopulations 1 and 2 and (ii) total number of cells and the proportion of slow-cycling cells (on the third day after treatments) when two doses of cell-cycle phase-specific chemotherapeutic drugs are given at times 500 hrs and 570 hrs (\~ 3 days apart). (a) G1 phase drug followed by G1 phase drug, (b) G1 phase drug followed by S-G2-M phase drug, (c) S-G2-M phase drug followed by S-G2-M phase drug (d) S-G2-M phase drug followed by G1 phase drug, (e) total number of cells and (f) percentage of cells in subpopulation 2 (slow-cycling cells).}
  \label{FigureR5}
\end{figure}
%%%%

\begin{figure}[t!]
  \begin{center}
      \includegraphics[scale=1.06]{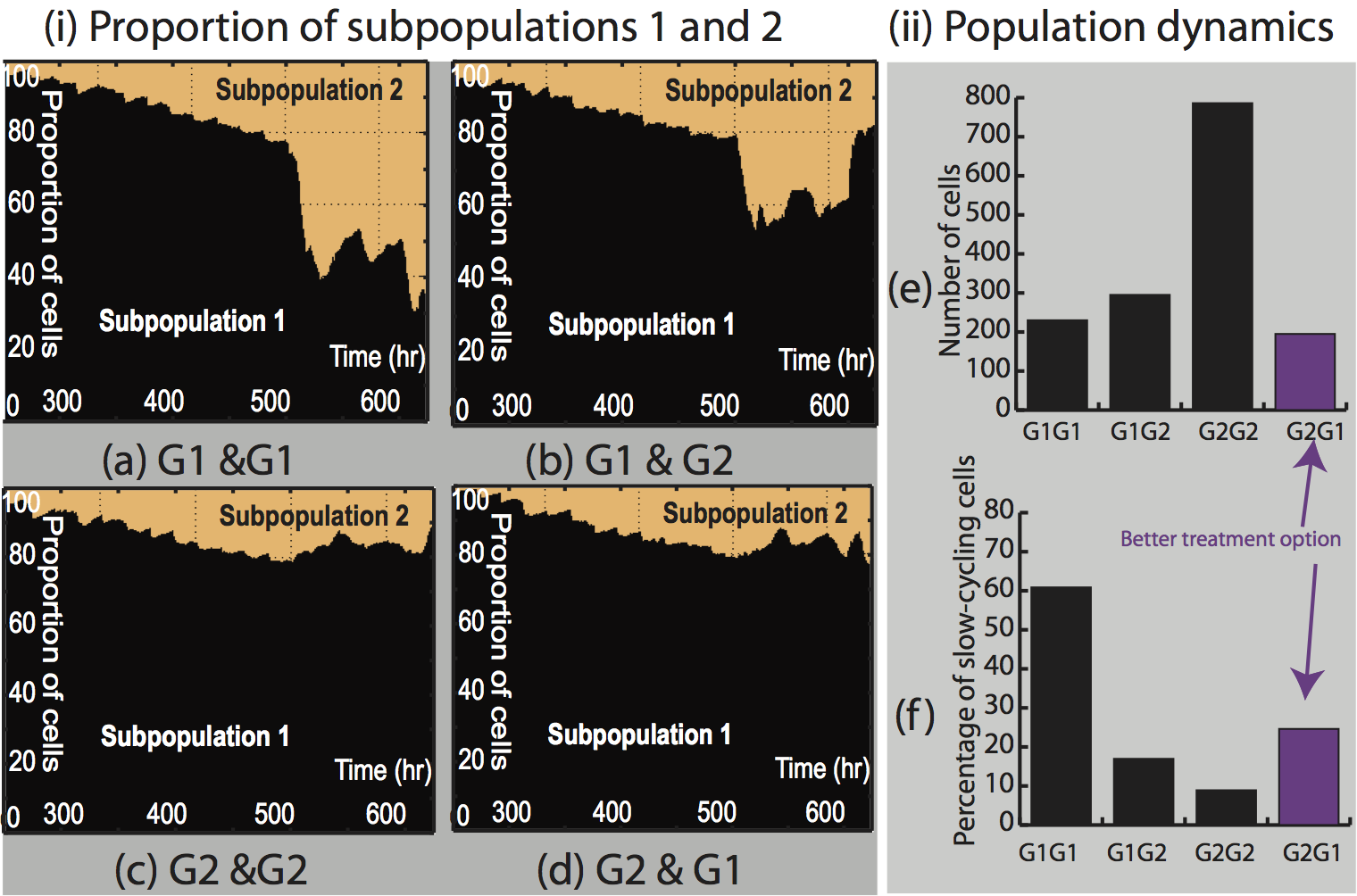}
      \end{center}
  \caption{Plots showing the (i) percentage of cancer cells in subpopulations 1 and 2 and (ii) total number of cells and the proportion of slow-cycling cells (on the third day after treatments) when two doses of cell-cycle phase-specific chemotherapeutic drugs are given at times 500 hrs and 620 hrs (\~ 5 days apart). (a) G1 phase drug followed by G1 phase drug, (b) G1 phase drug followed by S-G2-M phase drug, (c) S-G2-M phase drug followed by S-G2-M phase drug and (d) S-G2-M phase drug followed by G1 phase drug, (e) total number of cells and (f) percentage of cells in subpopulation 2 (slow-cycling cells).}
  \label{FigureR6}
\end{figure}
%%%%

\subsubsection*{Effects of chemotherapeutic drugs on a heterogeneous tumour: Better scheduling for a better outcome}	
Here we analyse the effects of multiple doses of cell-cycle phase-specific chemotherapeutic drugs on the heterogeneous tumour model by using two doses of drugs that act on either the G1 phase or the S-G2-M phases of the cell-cycle. The initial dose of the drug is delivered at time t = 500 hrs and the second dose of the drug is given either after 3 days or 5 days. The results are then analysed on the third day after the second dose (similar to the experimental protocol that used a single dose) to study the effects of the combined treatment and are plotted in Figures \ref{FigureR5} and \ref{FigureR6}.

Figure \ref{FigureR5}(i) shows the percentage of cancer cells in subpopulations 1 and 2 when the drugs are given with an interval of 3 days. We studied four different combinations of the phase-specific drugs, namely, (a) G1 phase drug followed by G1 phase drug, (b) G1 phase drug followed by S-G2-M phase drug, (c) S-G2-M phase drug followed by S-G2-M phase drug and (d) S-G2-M phase drug followed by G1 phase drug. The total number of tumour cells and the percentage of slowly-cycling cells on the third day after the second dose are shown in Figure \ref{FigureR5}(ii). These plots show that drug combinations with dose/doses of G1 phase-specific drug give a higher reduction in tumour mass than two doses of S-G2-M drugs. The combination of G1 \& G1 drugs killed the highest number of cancer cells, sparing the cancer cells that are in S-G2-M phase. Although this combination (G1 \& G1 drugs) resulted in higher cell-kill, it also enhances the enrichment of the slowly-cycling cancer cell subpopulation, making it a major subpopulation within the tumour mass. This is because the combination of G1 phase followed by G1 phase drug spares some of the slowly-cycling cancer cells that are in the extended G2/M phase as shown in the previous section and experiments \cite{Moore2012, Moore2012b, Moore2011, Schillert2012}. Figures \ref{FigureR5}(e) and \ref{FigureR5}(f) show that in this scenario, the worst possible combination is the delivery of two doses of S-G2-M drug while the combinations of G1 and S-G2-M drugs give comparatively better outcomes. Moreover, among the four combinations, the optimimal treatment option is S-G2-M drug followed by G1 drug, as it not only induces a higher cell-kill but also reduces the number of slowly-cycling tumour cells. Further analysis of cell-cycle dynamics (not shown) show that the introduction of either type of phase-specific drug alters the cumulative cell-cycle dynamics of the remaining cancer cells within the tumour mass for both subpopulations. While a G1-specific drug changes the cell-cycle dynamics of subpopulations 1 and 2 in favour of the next dose of G1 phase- and S-G2-M phase-specific drugs, respectively, an initial dose of S-G2-M phase-specific drug redistributes the cells in subpopulations 1 and 2 in favour of the G1 phase-specific drug. Hence, a dose of S-G2-M phase-specific drug followed by a G1 phase-specific drug induces a higher cell-kill than that of S-G2-M phase followed by S-G2-M phase combinations.

A similar analysis is performed when the doses are given with an interval of 5 days and the corresponding results are shown in Figure \ref{FigureR6}. These plots also show that a combination of two doses of G1 phase drug give the higher cell-kill, enriching the cells in the slowly-cycling subpopulation, while the delivery of two S-G2-M drug provides the minimal kill rate. Here, the best possible treatment option is proven to be the combination of G1 and S-G2-M phase-specific drugs, as similar to the above case. These simulation results suggest that the optimal chemotherapeutic drug delivery very much depends on the heterogeneity within and around a tumour mass and hence, patient-specific therapeutic scheduling and delivery is necessary to improve the clinical survival status. This also indicates the importance of adaptive therapy \cite{Gatemby2009}, since a higher cell-kill may not necessarily mean a better outcome.

\section*{Conclusions}

Although many tumours show an initial positive response to chemotherapeutic treatment, in some cases the long term response is not encouraging. A great hurdle in the successful treatment of various solid tumours is the incidence of tumour recurrence due to the development of drug resistance of one or more subpopulations of cancer cells. Currently there is no clear understanding of why cancers develop intrinsic drug resistance to several chemotherapeutic drugs and then recur after therapy. There are several reasons that may contribute to this drug-resistance of cancer cell populations such as variations in cell-cycle control, anti-apoptotic proteins, multi-drug resistance through the activation of cellular pumps and increased metabolic activities \cite{Ding2012, Navin2011, Koshkin2012, Moore2012, Moore2012b, Moore2011, Schillert2012}. Recent experimental studies \cite{Moore2012, Moore2012b, Moore2011, Schillert2012} have shown that slow-cycling subpopulations of cancer cells also play an important role in developing drug resistance. 

In this paper, we have presented a hybrid multiscale modelling approach (using a Compucell3D framework) to study the role of intrinsic tumour heterogeneity and slow-cycling subpopulations of cancer cells in cell-cycle mediated chemotherapeutic drug resistance. To achieve this, a slow-cycling subpopulation of cancer cells was introduced into a homogeneous tumour growth model through intracellular mutations in order to study the cytotoxic effects when two doses of cell-cycle-specific chemotherapeutic drugs were introduced. In general, the simulation results of the model re-confirmed the importance of patient-specific therapeutic protocols for the optimum delivery of anticancer therapies as illustrated in previous studies by Powathil et al. \cite{Powathil2012b, Powathil2013}. Moreover, the results from the heterogeneous model are qualitatively in very good agreement with the experimental results of \cite{Moore2012, Moore2012b, Moore2011} in showing that a slow-cycling subpopulation of cancer cells consists of a higher proportion of G2/M phase cells when compared with the rest of the cancer cells. Most importantly, this study also confirmed the hypothesis and experimental findings that a subpopulation of slow-cycling cancer cells exhibits some intrinsic cell-cycle-driven chemotherapeutic resistance, contributing towards tumour recurrence. 
This is an important factor, as the current chemotherapeutic drugs and radiotherapies target rapidly proliferating cancer cells that are in active phase of the cell-cycle, sparing some of the slow-cycling cancer cells, and ultimately leading to an enriched subpopulation of cancer cells with drug resistance. Our results indicate that cell-cycle-length-mediated chemotherapeutic resistance is mainly due to the fact that the currently available drugs  either target cancer cells in a particular phase/phases of the cell-cycle, thus sparing some of the cancer cells that are not in the targeted cell-cycle phase/phases. One effective way to address this issue is by delivering a chemotherapeutic drug that acts on all the cancer cells, irrespective of the cell-cycle phase or microenvironment status. However, since such a chemotherapeutic drug does not yet exist (to our knowledge), the next best option is the use of multiple doses of chemotherapeutic drugs that cumulatively act on all the cancer cells. Moreover, the insights from the simulation results suggest that a better understanding of cell-cycle-driven intracellular heterogeneity can play a greater role in the optimum delivery of cell-cycle-specific chemotherapeutic drugs, targeting the multiple subpopulations of cancer cells within a tumour mass more effectively.

\section*{Acknowledgments}
The authors gratefully acknowledge the support of the following :1) ERC Advanced Investigator Grant 227619, M5CGS: From Mutations to Metastases: Multiscale Mathematical Modelling of Cancer Growth and Spread, 2) NIH NIGMS R01 GM077138-04 "Competitive Renewal of Development, Improvement and Extension of the Tissue Simulation Toolkit - CompuCell3D".

\end{document}